\documentclass[preprint2]{emulateapj}
\usepackage{psfig}
\usepackage{graphicx}

\def\approxgt{\ifmmode \rlap{$>$}{}_{{}_{{}_{\textstyle\sim}}} \else%
$\rlap{$>$}{}_{{}_{{}_{\textstyle\sim}}}$\fi} 
\def\approxlt{\ifmmode \rlap{$<$}{}_{{}_{{}_{\textstyle\sim}}} \else%
$\rlap{$<$}{}_{{}_{{}_{\textstyle\sim}}}$\fi}
\def\farcm{\hbox{$.\mkern-4mu^\prime$}}
\def\farcs{\hbox{$.\!\!^{\prime\prime}$}}

\def\arcsec{\hbox{$^{\prime\prime}$}}

\def\src{IGR~J00291+5934}
\def\flx{erg cm$^{-2}$ s$^{-1}$}

\def\sax{SAX~J1808.4--3658}

\shorttitle{Observations of \src\ in quiescence}
\shortauthors{Jonker et al.}

\begin{document}

\title{Optical and X--ray observations of \src\ in quiescence}

\author{P.~G.~Jonker\altaffilmark{1,2}}
\affil{SRON, Netherlands Institute for Space Research, 3584~CA, Utrecht, The Netherlands}
\email{p.jonker@sron.nl}
\author{M.~A.~P.~Torres}
\affil{Harvard--Smithsonian  Center for Astrophysics, Cambridge, MA~02138, Massachusetts,
U.S.A.}
\email{mtorres@head.cfa.harvard.edu}
\author{D.~Steeghs\altaffilmark{1}}
\affil{Astronomy and Astrophysics, Department of Physics, University of Warwick, Coventry,  CV4~7AL, U.K.}
\email{d.t.h.steeghs@warwick.ac.uk}

\altaffiltext{1}{Harvard--Smithsonian  Center for Astrophysics, Cambridge, MA~02138, Massachusetts,
U.S.A.}
\altaffiltext{2}{Astronomical Institute, Utrecht University, 3508 TA, Utrecht, The Netherlands}

\begin{abstract}

We report on optical and X--ray observations of the accretion powered ms pulsar \src\ in
quiescence. Time resolved $I$--band photometry has been obtained with the 4.2~m William Herschel
Telescope, while a 3 ks {\it Chandra} observation provided contemporaneous X--ray coverage.
We found an unabsorbed 0.5--10 keV X--ray flux of 1$\times 10^{-13}$ \flx\ which implies that the source was in quiescence at the time of the optical observations. Nevertheless, the optical $I$--band light curve of \src\ shows evidence for strong flaring. After removal of the strongest flares, we find evidence for an orbital modulation in the phase folded $I$--band light curve. The overall modulation can be described by effects resulting from the presence of a superhump. Comparing our lightcurve with that reported recently we find evidence for a change in the quiescent base level. Similar changes have now been reported for 4 soft X--ray transients implying that they may be a common feature of such systems in quiescence. Furthermore, the maximum in our folded lightcurve occurs at a different phase than observed before. 

\end{abstract}



\section{Introduction}

Low--mass X--ray binaries (LMXBs) consist of a compact object, either a neutron
star or a black hole, that accretes from a late type companion star. The
companion star typically has a mass $\approxlt$ 1 M$_\odot$. Stellar evolution
theory predicts that neutron star LMXBs are the predecessors of the recycled millisecond
radio pulsars (\citealt{radsri1982}; \citealt{1982Natur.300..728A}). This link
between the neutron star LMXBs, their transient cousins, the X--ray
transients, and the millisecond radio pulsars has been established by the
discovery of the first accretion powered millisecond X--ray pulsar in 1998
(\citealt{1998Natur.394..344W}; \citealt{1998Natur.394..346C}). 

Subsequently, more of these systems have been found. At the time of writing there are ten
accretion powered millisecond X--ray pulsars known. In three of these sources pulsations are detected only intermittently (the transients HETE~J1900.1--2455, Aql~X--1 and SAX J1748.9-2021 in the globular cluster NGC~6440; \citealt{2007ApJ...654L..73G}, \citealt{2007arXiv0708.1110C}, \citealt{2007arXiv0708.1316A}). To date all of them are found in X--ray transients (see \citealt{2006tpr..conf...53W} for an observational overview). These transients flare--up during episodes of enhanced mass accretion rate onto the neutron star. It is thought that during outburst the enhanced mass accretion rate suppresses the radio pulsar mechanism precluding the detection of pulsed radio emission. Radio emission of these transients has been detected in outburst, but this is associated with synchrotron emission from a radio jet (e.g.~SAX~J1808.4--3658, \citealt{2002IAUC.7997....2R}; XTE~J0929--314 \citealt{2002IAUC.7893....2R}; IGR~J00921+5934, \citealt{2004ATel..355....1P}, \citealt{2004ATel..361....1F}). When the mass accretion rate stops/becomes very low the radio pulsar mechanism should turn on. However, despite deep searches using sensitive radio telescopes at epochs when one of these transients, SAX~J1808.4--3658, was in quiescence no radio millisecond pulsations have been found (\citealt{2003ApJ...589..902B}). The non--detection has been ascribed to the presence of absorbing material close to the binary.

In quiescence these systems are very faint in the optical. The counterpart can often not be
detected (e.g.~XTE~J0929--314 and XTE~J1814-331, the latter source was not detected down to
$R$=23.3; \citealt{2005ApJ...627..910K}). For the system XTE~J1751--305 the optical counterpart
was not discovered in outburst nor in quiescence (\citealt{2003MNRAS.344..201J}). There are two
noticeable exceptions;  SAX~J1808.4--3658 (\citealt{2001MNRAS.325.1471H}) and  IGR~J00291+5934
(\citealt{2004ATel..354....1F}; \citealt{2004ATel..356....1R}; \citealt{2004ATel..363....1S};
\citealt{2005ATel..395....1B}). For a comprehensive overview of the outburst and initial
quiescence observations of \src\ see \citet{2007astro.ph..1095T}. \citet{2007arXiv0707.3037D}
recently reported multi--band quiescent optical and near--infrared observations of \src. 

Using optical observations of SAX~J1808.4--3658 in quiescence, Homer et al.~(2001) found
evidence for a 9--15 per cent semi--amplitude modulation (the observed amplitude depends on the photometric band that is used for the observations). Those authors proposed that this would be due to X--ray irradiation from the neutron star. However, as pointed out by \citet{2003A&A...404L..43B}, assuming that SAX~J1808.4--3658 was indeed in quiescence at the time of the optical observations, the quiescent X--ray flux of SAX~J1808.4--3658 is too low by two orders of magnitude to explain the modulation in terms of X--ray heating. The absence of a double--humped morphology rules out that the modulation is due to ellipsoidal variations. \citet{2003A&A...404L..43B} proposed that the irradiation is caused by a turned--on radio pulsar instead of the quiescent X--ray emission. Evidence for a turned--on radio pulsar,  albeit indirect in this case (see \citealt{2004ApJ...614L..49C}), would reinforce the link between the LMXBs and the millisecond radio pulsars. Furthermore, a turned--on radio pulsar would have an important effect on the evolution of the mass--losing donor star, altering the evolutionary path of the binary (\citealt{1989ApJ...343..292R}). Due to the absence of pointed X--ray observations at the time of the optical observations of \citet{2001MNRAS.325.1471H} and \citet{2004ApJ...614L..49C}, low level X--ray activity could have remained unnoticed in X--rays but would heat the side of the companion star facing the neutron star, providing an alternative explanation for the observed optical modulations. However, simultaneous X--ray and optical observations of SAX~J1808.4--3658 reported by \citet{2007arXiv0710.1552H} forego this possibility.

In this Manuscript, we present phase resolved photometric observations of the optical
counterpart of \src\ in quiescence obtained with the 4.2~m William Herschel Telescope (WHT).
In addition we present our analysis of a short contemporaneous {\it Chandra} observation of
\src\ obtained with the aim to determine the contemporaneous X--ray flux. 

\section{Observations, analysis, and results}

\subsection{{\it Chandra} X--ray observations}

We have observed \src\ with the back--illuminated S3 CCD--chip of the Advanced CCD Imaging
Spectrometer (ACIS) detector on board the {\it Chandra} satellite. The observation started
on Sept.~13, 2006 at 13:33 (UTC; MJD 53991). The data telemetry mode was set to {\sl very faint} to
allow for a thorough background subtraction. Due to windowing of the ACIS--S CCD a frame
time of 0.4104 s has been used. We have reprocessed and analysed the data using the {\it
CIAO 3.4} software developed by the Chandra X--ray Center using CALDB version 3.3.0.1 to
take full advantage of the {\sl very faint} data mode. In our analysis we have selected
events only if their energy falls in the 0.3--7 keV range. The 0.3--7 keV background count
rate was always lower than 0.4 counts s$^{-1}$ during our observation. The net on--source
exposure time is 2.88 ks. 

Using {\sl wavdetect} we detected two sources CXC~J002911+593420 and \src. We detect 22 photons
from \src\ during the 2.88 ks observation, leading to a source count rate of 7.6$\times 10^{-3}$
counts s$^{-1}$. Using {\sl xspec} version 11.3.2p (\citealt{ar1996}) we have fitted the
spectrum of \src\ using Cash statistics (\citealt{1979ApJ...228..939C}) to an absorbed powerlaw
model. We held fixed the interstellar extinction to the value of 4.6$\times 10^{21}$ cm$^{-2}$
favored by Torres et al.~(2007) and the powerlaw index to 2. We derive an absorbed 0.5--10 keV
source flux of 7$\times 10^{-14}$ \flx\ and an unabsorbed 0.5--10 keV flux of 1$\times 10^{-13}$
\flx. The source flux is consistent with that derived previously by \citet{2005MNRAS.361..511J} and Torres et al.~(2007). We searched for variability in the rate of arrival of the photons but
we found none. A Kolmogorov--Smirnov (\citealt{prteve1992}) test showed that the probability
that the data are consistent with the null--hypothesis of a constant photon arrival rate is 63\%. We conclude that the source was in quiescence during our optical observations.

\subsection{WHT optical observations}

We obtained Harris $I$--band images using the Auxiliary Port Imager (AUX) instrument mounted on
the 4.2~m WHT telescope at the Roque de Los Muchachos Observatory, La Palma, Spain. On September
13 and 14, 2006 (MJD 53991 and 53992 UTC) observations with an exposure time of 600~s, totalling
10.7 hours of data, were obtained. The observing conditions were good with a photometric sky and
a seeing varying between 0\farcs7--1\farcs6 and between 0\farcs7--1\farcs0 and with a mean
seeing of 1\farcs0 and 0\farcs74 on Sept.~13 and Sept.~14, respectively. Given the faintness of the quiescent counterpart we have rejected images with seeing $\geq$1\arcsec. This left us with 24 images for Sept.~13 and 40 for Sept.~14. We used a 2 $\times$ 2 on--chip binning, therefore AUX delivered a circular field of view with a diameter of 1\farcm8 sampled at 0\farcs22/pixel. Frames were debiased and then flatfielded using dome flatfield observations.

\begin{figure} \epsscale{1.25} \plotone{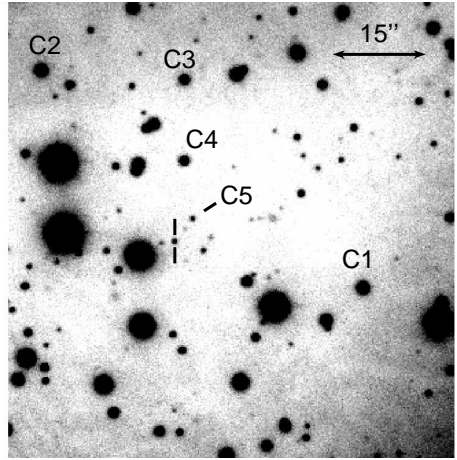} \caption{Optical $I$--band finder chart of \src. North is up East to the left, the scale of the image is indicated in the top--right part of the figure. The optical counterpart to \src\, is indicated
with two vertical lines. The stars labeled C1 to C4 are comparison
stars with $I$--band magnitudes $18.41 \pm 0.02$, $18.55 \pm 0.03$, $19.34
\pm 0.05$ and $19.41 \pm 0.03$, respectively. The star marked with C5
is a nearby comparison star of similar brightness as \src\ ($I = 21.80 \pm 0.05$).
\label{finder}} \end{figure}

We applied both aperture photometry and point--spread function fitting
on each of the 600~s images using DAOPHOT in \textsc{IRAF}\footnote{\textsc {iraf} is distributed by the National Optical Astronomy
Observatories} to compute the instrumental magnitudes of the detected stars. The results obtained with both methods are
consistent with each other. Flux calibration of the field was
performed by observing two Landolt standard stars and differential
photometry was used to derive the source flux variability as a
function of time. The photometric results given here are with respect to the four field
stars shown in Figure~\ref{finder}. These are the brightest isolated stars
available in the unvignetted field of view of AUX that were recorded
in the linear regime of the CCD.

\begin{figure} \includegraphics[scale=.37,angle=270]{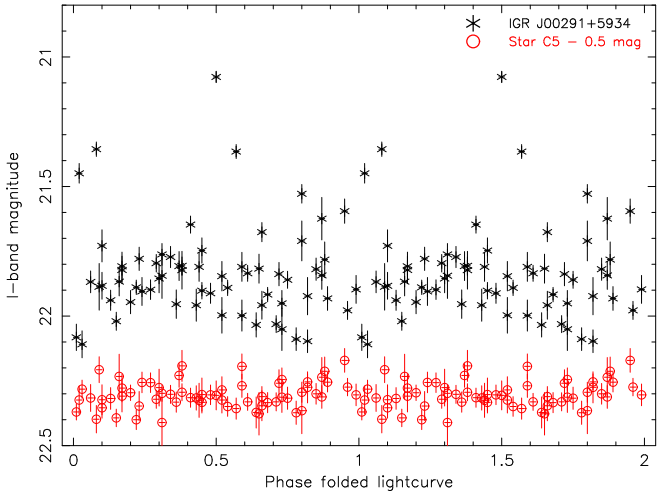}
\caption{Optical $I$--band folded light curve of \src\ (stars) obtained with the 4.2~m
WHT on Sept.~13 and 14, 2006. Strong flaring of \src\ can be seen. The strong flaring precludes the detection of clear trends in the $I$--band magnitude
as a function of orbital phase. The open circles shows the $I$--band magnitude of a nearby star of
comparable brightness (star C5 in Figure~\ref{finder}) measured simultaneously. The points for star C5 have been scaled down by 0.5 magnitude for display purposes. Phase zero corresponds to the epoch of 90$^\circ$ mean longitude as given by Galloway et al.~(2005).
\label{optlc}} \end{figure}

\begin{figure} \includegraphics[scale=.45,angle=0]{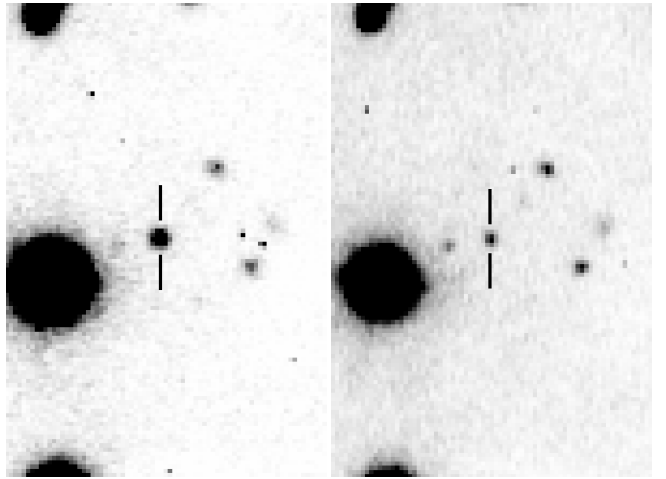} \caption{The image corresponding to the strongest flare visible in Figure ~\ref{optlc} is displayed on the {\it left}. On the {\it right} is the image taken 20 minutes after the flare. North is up East to the left. \label{imflare}}
\end{figure}

We plot the $I$--band magnitude of \src\ (crosses) and those of a nearby comparison star of similar brightness (open circles; star C5) as observed on Sept.~13 and 14, 2006 in Figure~\ref{optlc} folded on the orbital period of \src. We have used the pulsar ephemeris of \citet{2005ApJ...622L..45G}. The error in propagating the orbital ephemeris to the time of our observations is $\approx$0.01 in phase. Phase zero is superior conjunction of the neutron star
(i.e.~the epoch of 90$^\circ$ mean longitude). The brightness of the comparison star (star C5 on Figure~\ref{finder}) is consistent with being constant. In contrast, the folded lightcurve of \src\ displays several large flares (see also Figure~\ref{imflare}). The average $I$--band magnitude of \src\ and the variance therein is 21.83 and 0.18 magnitudes, respectively.

\begin{figure} \includegraphics[scale=.37,width=8cm,height=8cm,angle=0]{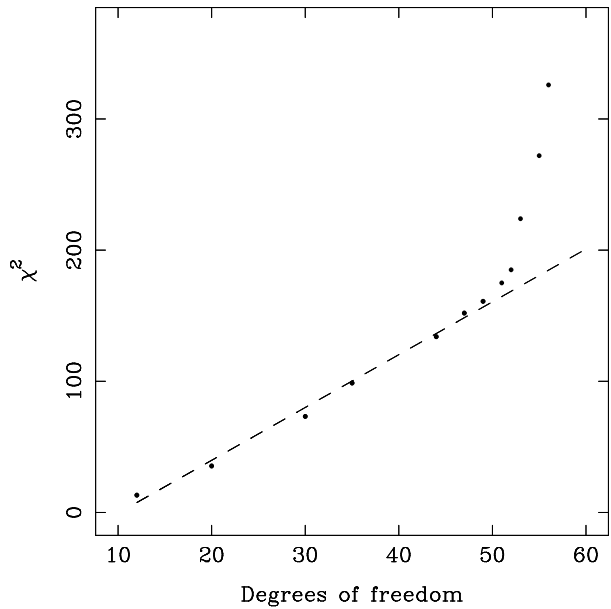} \caption{The $\chi^2$ of the fit of a constant plus sinusoid to the phase folded light curve of \src\ plotted against the number of degrees of freedom (d.o.f.) in the fit for the various magnitude thresholds that we investigated (see  Table~\ref{tbl1}). The drawn line is the linear correlation described by the data points below d.o.f.=50. The transition between the linear correlation and a more steep correlation occurs around magnitude 21.75. \label{flaredef}} \end{figure}

In an attempt to determine whether there are variations in the quiescent optical brightness due
to a change in aspect of the Roche--lobe filling companion star as a function of the orbital
phase, we have removed flares from the lightcurve of both nights.
In order to come to a definition of a flare we have varied the magnitude threshold above which we identify a data point as a flare. Next, for a range of potential magnitude thresholds we fold and average the data in 10 orbital phase bins and we determine the $\chi^2$ and the number of degrees of freedom (d.o.f.) of a fit of a sinusoid plus a constant to the folded data. We show the resuls of this in Figure~\ref{flaredef}. It is clear that the  $\chi^2$ strongly decreases when more and more stringent magnitude thresholds are taken. The reduction in $\chi^2$ levels off around a threshold magnitude of 21.75 (see Table~\ref{tbl1}). For more stringent magnitude limits the decrease in $\chi^2$ is proportional to the reduction in the number of degrees of freedom as indicated by the straight dashed line in Figure~\ref{flaredef}. Such behavior is expected when clipping data points that are not flares.

\begin{deluxetable}{cccc}
\tabletypesize{\scriptsize}
%\rotate
\tablecaption{The effect of the definition of a flare on a fit of a constant plus sinusoid to the folded lightcurve. \label{tbl1}}
\tablewidth{0pt}
\tablehead{
\colhead{Magnitude limit} & \colhead{${\rm \chi^2_{red}}$} & \colhead{${\rm \chi^2}$} & \colhead{D.o.f.}}
\startdata
21.95	& 1.1	& 13.2	& 12\\
21.90	& 1.77	&  35.4	& 20\\
21.85	& 2.44	& 73.2	& 30\\
21.825	& 2.82	& 98.7	& 35\\
21.80	& 3.05	& 134	& 44\\
21.775	& 3.23	& 152	& 47\\
21.75	& 3.29	& 161	& 49\\
21.725	& 3.43	& 175	& 51\\
21.70	& 3.56	& 185  &     52\\
21.65	& 4.23	& 224	& 53\\
21.60	& 4.95	& 272	& 55\\
21.55	& 5.82	& 326	& 56\\
\enddata
%\tablenotetext{a}{A}
\end{deluxetable}

\begin{figure} \includegraphics[scale=.37,width=8cm,height=8cm,angle=0]{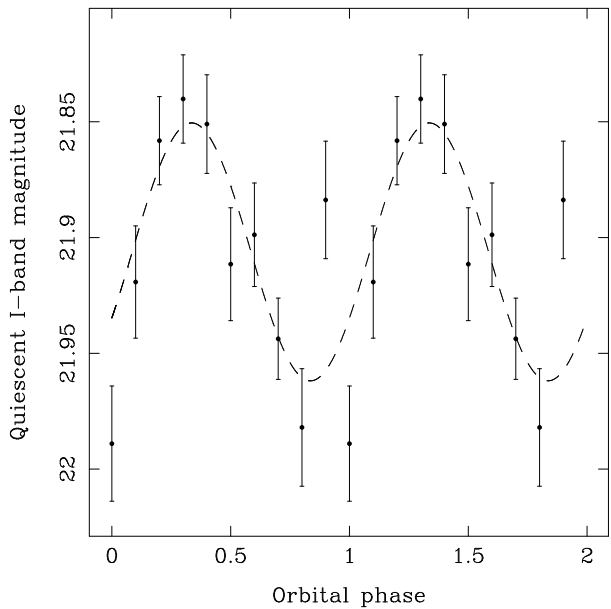} \caption{Phase
folded optical $I$--band light curve of \src\ after excluding flares in the light
curve (i.e.~data points with $I<$21.75). The dashed line is the best--fitting model of a sinusoid plus a constant. The sinusoid reaches a maximum at phase 0.34 $\pm$ 0.03. For
clarity two orbital cycles are plotted. \label{fold}} \end{figure}

Using 21.75 as our magnitude threshold above which we define data points as flares, the average $I$--band magnitude and the variance therein becomes 21.90 and 0.09 magnitudes, respectively. The variance is slightly higher than $\sim$0.05 which is found for several stars of similar brightness that do not vary, implying that variability due to intrinsic variations is still present in the lightcurve of \src. The resultant folded light curve is shown in Figure~\ref{fold}. The most striking feature is 
the presence of a clear orbital modulation. In order to quantify the modulation we have fit the folded lightcurve with a fit function consisting of a constant plus a sinusoid. An F--test gives that the improvement of a fit of a constant plus sinusoid with respect to a constant has a probability of 2\% to occur due to chance. We derive $(6\pm1)\times10^{-2}$ magnitudes and 0.34$\pm$ 0.03 for the semi--amplitude and phase of the sinusoid, respectively. The value of the constant is 21.91$\pm$0.01 magnitudes. In addition to the sinusoidal variation there is suggestive evidence for an increase in source brightness at phase 0.9. In principle, an alignment of small flares in the phase bin centered on phase 0.9 can have caused an increase above the sinusoidal modulation. Ellipsoidal modulations due to the Roche--lobe filling companion star are not detected.

\section{Discussion}

Comparing the source X--ray flux during a 3 ks--long {\it Chandra} observation of \src\ with that observed on previous occasions when the source was in quiescence, we conclude that \src\ was in quiescence on Sept.~13, 2006. Optical $I$--band observations acquired with the 4.2~m WHT telescope on Sept.~13 and 14, 2006 reveal strong variability with flares of $\sim$1 magnitude on timescales of tens of minutes. After removal of flares we have phase folded our data. A clear sinusoidal variation is present. In our data set the orbital phase at which the source brightness is maximal is 0.34 $\pm$ 0.03. \citet{2007arXiv0707.3037D} report a maximum consistent with phase 0.5. Furthermore, their average $I$--band magnitude of 22.4$\pm$0.2 for \src\ is 0.50$\pm$0.22 lower than our average magnitude (after removal of flares larger than $I$=21.75). Hence, the brightening is only significant at the 2.3~$\sigma$ level and \citet{2007arXiv0707.3037D} obtained the $I$--band observations under non--photometric conditions. Interestingly, \citet{2007arXiv0707.3037D} do not report evidence for flaring. The lack of flaring together with the lower average brightness suggests that the source showed activity at the time of our optical observations above that found by \citet{2007arXiv0707.3037D}. 

In order to investigate the nature of the flares and the observed variability in the phase folded optical light curve, we compared the quiescent properties of \src\ observed by us with those reported by Torres et al.~(2007) and \citet{2007arXiv0707.3037D} and with those reported for several other sources, most notably Cen~X--4 (\citealt{1989A&A...210..114C}, \citealt{1993MNRAS.265..655S}). Optical flares similar to those observed in \src\ in quiescence have been reported for several sources (e.g.~\citealt{1989A&A...210..114C}, \citealt{2003ApJ...582..369Z}, \citealt{2003MNRAS.340..447H}). However, flares as large as the largest that we observe in \src\ have not been reported to our knowledge. In addition, we note that a variation of about half a magnitude in the average brightness similar to that mentioned above for \src\ has been reported for Cen~X--4, XTE~J2123--058 and the black hole X--ray transient A~0620--00 (\citealt{1989A&A...210..114C}, \citealt{2004ApJ...610..933T}, \citealt{2008ApJ...673L.159C}, respectively).

Strong flares of similar duration (tens of minutes) have been observed in mid and late M dwarfs
(e.g.~see \citealt{2006MNRAS.367..407R} and references therein). However, \citet{2003MNRAS.340..447H} showed that the flares in quiescent LMXBs are associated with the accretion disk. Indeed, the combination of strong flaring and the lack of clear ellipsiodal variations in the phase folded light curve (this work as well as \citealt{2007arXiv0707.3037D}), imply that the quiescent light is not dominated by a Roche lobe filling, but otherwise unperturbed donor star. If the flares are from the donor star their amplitude must be larger still. Instead, (superhumps from) a residual accretion disk, the accretion stream, and/or effects from an irradiated donor star are potentially important contributors to the quiescent $I$--band light. Torres et al.~(2007) derive a similar conclusion on the basis of the quiescent intrinsic $R-K$ color and constraints on the donor star from the pulsar mass function.  

The absence of ellipsoidal variations in the phase folded light curves of both \src\ as well as \sax\ is consistent with the very low contribution of a (non--irradiated) brown dwarf to the optical light. Overall, the phase folded quiescent $I$--band light curve of \src\ that we find resembles that of the average quiescent $V$--band light curve of Cen~X--4 observed by \citet{1989A&A...210..114C} in 1984--1988 (see their figure 7). \citet{1989A&A...210..114C} show that a model where a large fraction of the optical light comes from the heated hemi--sphere of the companion star or an accretion disk attenuated by electron scattering in an accretion wake can describe the data well. In particular such a model can explain the lower brightness at orbital phases 0.5--0.7. The heating of the companion star in ms X--ray pulsars in quiescence has been ascribed to the turn--on of an active radio pulsar (\citealt{2003A&A...404L..43B}). Such an effect might be underlaying the folded lightcurve of \src, however, due to the (likely) higher level of activity during our observations additional effects seem to be important. An important clue comes from the phase difference between the maximum of the sinusoidal modulation reported by \citet{2007arXiv0707.3037D} and that reported in this work. This suggests the presence of a modulation that has a period unequal to the orbital period, such as a superhump. Potentially, the overall morphology of our folded lightcurve can indeed be described by effects originating in a superhump. Superhumps have been detected in the black hole candidate XTE~J1118+480 while near quiescence (\citealt{2002MNRAS.333..791Z}). As explained in \citet{1991MNRAS.249...25W}, superhumps can occur in systems when the ratio between the companion star and the compact object mass is $\approxlt$ 0.3. An accretion stream that perturbs the outer disk facilitates the excitation of tidal torques that are setting off the superhump resonance (\citealt{1991MNRAS.249...25W}). The hint for a brightening of the source around phase 0.9 visible in Figure~\ref{fold} can be explained by the presence of an accretion stream and hot--spot while the mass ratio is very likely (much) lower than 0.3. However, from studies of Cataclysmic Variables it is known that the superhump only develops during outburst, most likely since the mass and hence angular momentum transfer rate through the disk in quiescence is too low to allow for a large extent of the disk. Hence, it is not clear whether all the conditions necessary for the occurrence of superhumps are fullfilled in \src\ during our observations. Interestingly, \citet{2007arXiv0710.3202N} also found evidence for the presence of an eccentric, precessing disk in the black hole candidate A~0620-000 in quiescence. As superhumps can potentially explain the overall variability observed in the optical in several systems while in quiescence, further investigation is necessary to test whether during phases of (enhanced) mass transfer in quiescence a superhump can develop in quiescent LMXBs with mass ratios less than 0.3.

\acknowledgments

PGJ acknowledges support from NASA grant G06-7032X and from the Netherlands
Organisation for Scientific Research. DS acknowledges an STFC Advanced Fellowship. We would like to thank the referee for his/her comments that helped improve the paper. {\it Facilities:} \facility{William Herschel
Telescope}, \facility{CXO (ACIS)}.

\end{document}